\documentclass{aastex631}

\usepackage{color,soul}
\usepackage{comment}

\shorttitle{Differential Flow of the Young solar wind }
\shortauthors{Mostafavi et al.}

\graphicspath{{./}{figures/}}

\begin{document}

\title{Alpha-proton Differential Flow of the Young Solar Wind: Parker Solar Probe Observations}


\author[0000-0002-3808-3580]{P. Mostafavi}
\affiliation{Johns Hopkins Applied Physics Laboratory, Laurel, MD 20723, USA}
\email{parisa.mostafavi@jhuapl.edu}

\author[0000-0003-2079-5683]{R. C. Allen}
\affiliation{Johns Hopkins Applied Physics Laboratory, Laurel, MD 20723, USA}

\author{M. D. McManus}
\affiliation{Space Sciences Laboratory, University of California, Berkeley, CA 94720, USA}

\author[0000-0003-1093-2066]{G. C. Ho}
\affiliation{Johns Hopkins Applied Physics Laboratory, Laurel, MD 20723, USA}

\author[0000-0003-2409-3742]{N. E. Raouafi}
\affiliation{Johns Hopkins Applied Physics Laboratory, Laurel, MD 20723, USA}

\author[0000-0001-5030-6030]{D. E. Larson}
\affiliation{Space Sciences Laboratory, University of California, Berkeley, CA 94720, USA}

\author[0000-0002-7077-930X]{J. C. Kasper}
\affiliation{Climate and Space Sciences and Engineering, University of Michigan, Ann Arbor, MI 48109, USA}

\author[0000-0002-1989-3596]{S. D. Bale}
\affiliation{Space Sciences Laboratory, University of California, Berkeley, CA 94720, USA}

\begin{abstract}
The velocity of alpha particles relative to protons can vary depending on the solar wind type and distance from the Sun \citep{Marsch_2012_review}. Measurements from the previous spacecraft provided the alpha-proton's differential velocities down to 0.3 au. Parker Solar Probe (PSP) now enables insights into differential flows of newly accelerated solar wind closer to the Sun for the first time. Here, we study the difference between proton and alpha bulk velocities near PSP perihelia of Encounters 3-7 when the core solar wind is in the field of view of the Solar Probe Analyzer for Ions (SPAN-I) instrument. As previously reported at larger heliospheric distances, the alpha-proton differential speed observed by PSP is greater for fast wind than the slow solar wind. We compare PSP observations with various spacecraft measurements and present the radial and temporal evolution of the alpha-proton differential speed. The differential flow decreases as the solar wind propagates from the Sun, consistent with previous observations. While Helios showed a small radial dependence of differential flow for the slow solar wind, PSP clearly showed this dependency for the young slow solar wind down to 0.09 au. 
Our analysis shows that the alpha-proton differential speed's magnitude is mainly below the local Alfv\'en speed. Moreover, alpha particles usually move faster than protons close to the Sun.
PSP crossed the Alfv\'en surface during its eighth Encounter and may cross it in future Encounters, enabling us to investigate the differential flow very close to the solar wind acceleration source region for the first time.  


\end{abstract}


\section{Introduction} \label{sec:intro}
The two primary ion components of the solar wind are protons and alpha particles (fully ionized Helium), which have different properties from one another, such as density, velocity, and temperature \citep{Marsch_2012_review, Verscharen_etal_2019_review}. 
Alpha particles, which play essential role in understanding the fundamental nature of solar wind, have a much smaller number density compared to protons \citep{Bame_etal_1977}. 
Furthermore, their differential streaming with respect to protons was observed for the first time by \cite{Robbins_etal_1970_He}. Since then, the alpha-proton differential velocity has been investigated by analyzing different spacecraft observations at various heliospheric locations. Even though alpha particles are heavier than protons, they tend to be faster than protons at distances close to the Sun \citep{Feldman_Marsch_1997}. Several works proposed different mechanisms for the super-acceleration and heating of alphas including drift instabilities \citep{Verscharen_etal_2013_AIC_instability}, resonant absorption of ion-cyclotron waves \citep{Isenberg_Vasquez_2009_heating_He_corona,Kasper_etal_2013_heating_He_IC}, and low-frequency Alfv\'en wave turbulence \citep{Chandran_2010_heating_He_corona}.

 Previous spacecraft observations showed the dependence of the alpha-proton differential streaming on the solar wind speed \citep{Marsch_etal_1982_He, Steinberg_etal_1996, Durovcov_etal_2017_He}. The slow solar wind, thought to be associated with the streamer belt, typically comprises protons and alphas with comparable velocities.  As the solar wind increases in speed up to that of the high-speed streams, likely originating from open magnetic fields associated with coronal holes, the alpha-proton differential streaming similarly increases. 
 The dependence of the differential flow on the solar wind speed is more pronounced near the Sun and decreases with increasing heliodistance \citep[see Fig. 10 of ][]{Marsch_etal_1982_He}.
Thus, the dependence of the alpha-proton differential speed on the solar wind speed may indicate differences in solar wind acceleration between these different regions.

Various spacecraft at different heliographic distances enabled the study of the radial evolution of solar wind protons, alpha particles, and their differential velocities in the inner heliosphere from $\sim$0.3 to 5.5 au. Helios made observations from 0.3 to 1 au and showed that the average differential flow decreases with increasing heliocentric radial distance \citep{Marsch_etal_1982_He, Durovcova_etal_2019_Helios}.
\cite{Durovcov_etal_2017_He} analyzed Wind observations at 1 au over 20 years and, by comparing with Helios data, confirmed the differential velocity's radial dependence. 
Furthermore, the alpha-proton differential speed at distances beyond Earth measured by Ulysses from 1 to 5.4 au has been previously reported  \citep{Neugebauer_etal_1996, Reisenfeld_etal_2001_He_Ulysses}. The observations showed that the differential speed continued to decrease with increasing distance until no significant differences were detected at large distances from the Sun ($\sim$4 au). 

Additionally, observations of the solar wind by previous spacecrafts show that the absolute value of the differential flow between alphas and protons usually stays lower than, or comparable to, the local Alfv\'en speed \citep{Marsch_etal_1982_He, Neugebauer_etal_1996, Berger_etal_2011_he, Durovcov_etal_2017_He}. 
The proposed explanation for the velocity limitation of the differential velocity is the wave-particle interaction driven by alpha-proton instabilities, which reduces the differential velocity down to the Alfv\'en speed \citep[more details in][]{Gary_etal_2000_instablity, Verscharen_etal_2013_AIC}. 
The alpha-proton differential velocity ($V_{\alpha p}$) at about 0.3 au is often near this upper limit (i.e., the local Alfv\'en speed, $V_A$) \citep{Marsch_etal_1982_He, Durovcov_etal_2017_He}.
\cite{Neugebauer_etal_1996} showed that the $V_{\alpha p}/V_A$ decreases as the solar wind propagates and becomes close to zero at distances greater than $\sim$4 au, as observed by Ulysses.  One of the suggested mechanisms that regulate the differential flow with increasing distance is Coulomb collisions \citep{Kasper_etal_2008_he, Kasper_etal_2017_preferential_heating}. Collisional age, which is the ratio of the time between alpha-proton Coulomb collisions to the transit time of the solar wind, increases with distance from the Sun and thus decreases the differential flow by leading to the solar wind to thermodynamic equilibrium.

The present paper focuses on the alpha-proton differential flow of the young solar wind near the Sun (within 0.3 au) for the first time. In section \ref{sec:psp}, we show the selection of the Parker Solar Probe (PSP) data.  We statistically analyze PSP observations during close approaches of Encounters 3-7 down to $\sim$0.1 au from the Sun to study this differential flow (section \ref{sec:comp}). 
Moreover, we compare PSP measurements with observations from Helios, Wind, and Ulysses, spanning a wide range of heliospheric distances. This comparison shows the radial and temporal evolutions of the proton, alpha, and their relative speed. 
We compute the ratio between the alpha-proton differential speed and the local Alfv\'en speed of different solar wind types (i.e., slow, intermediate, and fast winds) to compare with previous studies. 
These results expand our knowledge about the solar wind's origin and evolution from the Sun to larger distances. In section 4, we summarize and conclude the paper.

\section{Data Sources and Selection} \label{sec:psp}
PSP, launched in 2018, is a mission to study our Sun and young solar wind ions closer than any spacecraft before \citep{Fox_etal_2016}. 
The Solar Wind Electrons Alphas and Protons (SWEAP) on PSP consists of two ion sensors, the Solar Probe Cup (SPC) and the Solar Probe Analyzer for Ions (SPAN-I), that combined provides a continuous view of bulk solar wind ions \citep{Kasper_etal_2016_SWEAP_instrumet}. 
SPC is a Faraday cup that points directly to the Sun while
SPAN-I consists of a ram-facing electrostatic analyzer and Time-of-Flight section, which enables the differentiation of ions with distinct masses by sorting particles by their mass-per-charge ratio 
\citep[more details in][]{Livi_etal_2021_Span_i}. 
Due to the increasing spacecraft lateral velocity closer to the Sun, the solar wind ion flow in the spacecraft frame appears to move into the ram-facing side of PSP and consequently is within the field of view of SPAN-I. 
This paper, which uses the SPAN-I instrument to study alpha particles and their relative velocities to protons, focuses solely on intervals during the closest approaches of PSP when the tangential velocity of the spacecraft is sufficient to enable SPAN-I observations of the core of the particle distribution.
Since the solar wind was largely not in the field of view of SPAN-I for the first two perihelia, we limit our statistical analysis to Encounters 3 through 7 in this paper. While obtaining velocities from the partial moments \citep{Livi_etal_2020a, Livi_etal_2020b} during times when the core of the distribution is within the field-of-view of SPAN-I is generally robust, densities derived from these partial moments may still not be fully accurate, as the full ion distribution is not captured in the measurements. Therefore, we obtain the density from the analysis of the plasma quasi-thermal noise (QTN) spectrum measured by the PSP/FIELDS Radio Frequency Spectrometer (RFS) \citep[more details in][]{Moncuquet_etal_2020_QTN_PSP}. 
During these perihelia, PSP spacecraft had a perihelia of 0.166 au (35.67 $R_s$) during Encounter 3 and 0.095 au (20.35 $R_s$) during Encounter 7. 
We use the SPAN-I data product ``sf0a" for the differential energy flux of alphas with the proton contamination subtracted, along with the ``sf00" data product, which indicates differential energy flux of protons.
The time resolutions of the SPAN-I proton and alpha particle observations are 7~s and 14~s, respectively. 
We also use magnetic field data from the FIELDS instrument \citep{Bale_etal_2016_FIEDS_PSP, Bale_etal_2020} to compute the local Alfv\'en speed as
\begin{equation}
    V_A = B/\sqrt{\mu_0 (n_p m_p + n_\alpha m_\alpha)} = \tau B/\sqrt{\mu_0 n_e m_p},
\end{equation}
where B is the magnetic field, $\mu_0$ is the permeability of vacuum, and $m_{p}$ and $m_{\alpha}$ are the proton and alpha masses. $n_p$, $n_\alpha$, and $n_e$ are the proton, alpha particle, and electron number densities, respectively.  Due to the before-mentioned uncertainty in the density computed from SPAN-I partial moments, we compute the Alfv\'en speed using the $n_e$. $\tau$ is a multiplicative factor proportional to the relative abundance of alpha particles in the solar wind. Using a range of alpha particle abundances from $2\%$ to $6\%$ of the solar wind, $\tau$ ranges from 0.98 to 0.95, respectively. Thus, the Alfv\'en speeds reported in this paper (which assumes $\tau$ = 1) may overestimate the Alfv\'en speed by a $2\%$  to $5\%$.

The paper focuses on statistical analysis of the differential velocity between protons and alpha particles. 
Some of the previous works analyzing alpha-proton differential velocity simply used the magnitude of the proton and alpha velocities to calculate their velocity differences $|V_\alpha| - |V_p|$ \citep[e.g.,][]{Neugebauer_etal_1994, Reisenfeld_etal_2001_He_Ulysses, Kasper_etal_2008_he}. In this work, however, we calculate the differential speed based on $|V_{\alpha p}| = |V_\alpha - V_p|$ to consider the directions of both vectors \citep[e.g., similar to work of][]{Steinberg_etal_1996,Berger_etal_2011_he, Durovcov_etal_2017_He}. 
Furthermore, the cases with alpha particles faster than protons and vice versa are calculated using
\begin{equation}
    V_{\alpha p } = sign(|V_{\alpha}| - |V_p|) |V_{\alpha} - V_p|. 
\end{equation}

\section{Observations and discussions} \label{sec:comp}
\begin{figure}[ht!]
\plotone{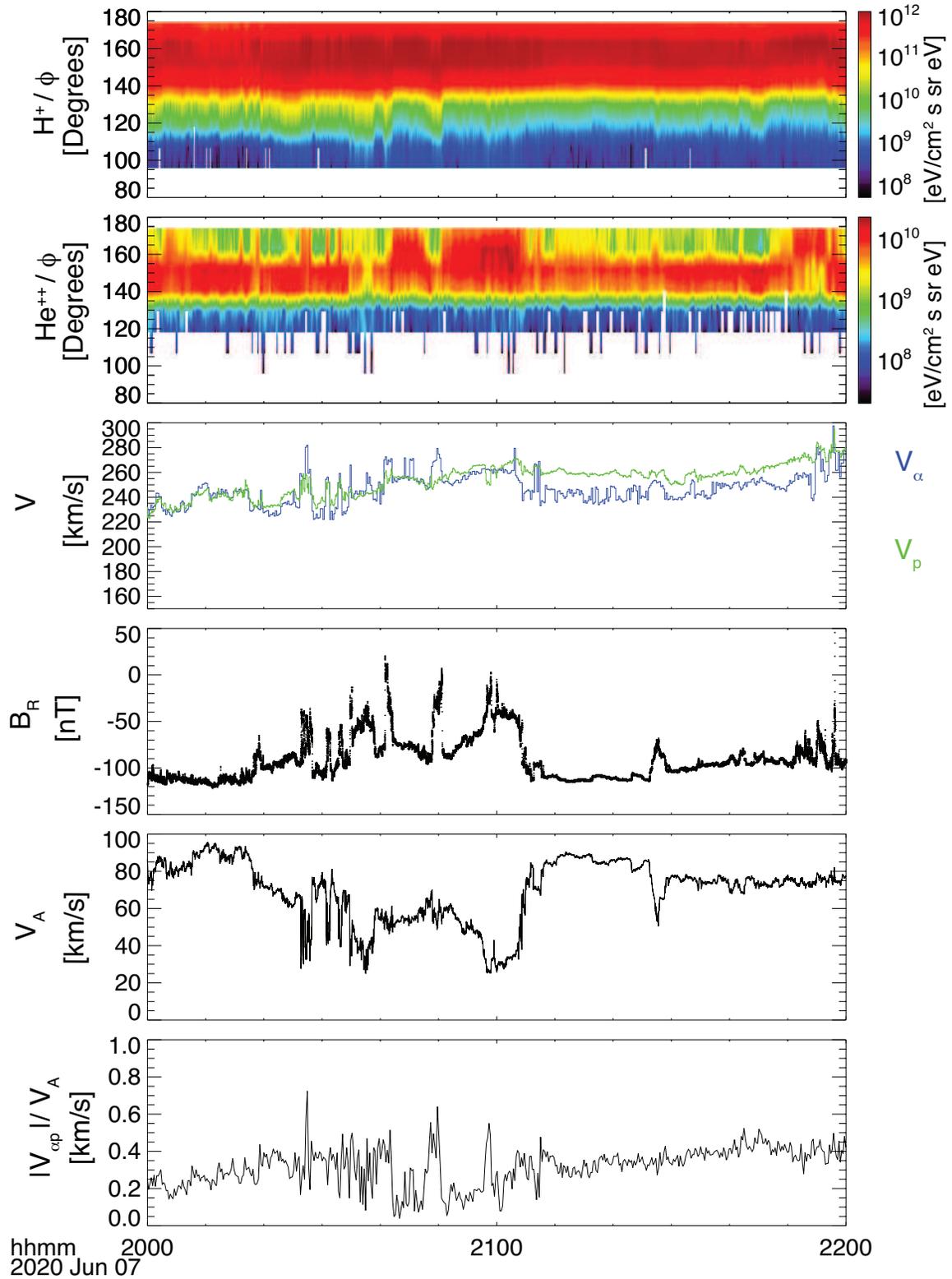}
\caption{Solar wind properties over two hours interval during Encounter 5. From top to bottom are  $H^+$ and $He^{++}$ flux as a function of azimuthal angle ($\phi$) in the spacecraft frame, proton and alpha velocities, the radial component of the magnetic field, Alfv{\'e}n speed, and magnitude of differential speed over local Alfv{\'e}n speed. \label{plasma_properties}}
\end{figure}

Figure \ref{plasma_properties} displays an example of the solar wind properties during a two-hour interval from Encounter 5.  The top two panels show the azimuthal H$^+$ and He$^{++}$ fluxes (in the spacecraft frame) over time, respectively, illustrating the variation of the peak of the plasma distribution within the field of view of SPAN-I. In order to have reliable plasma moments for SPAN-I, we only investigate times when the core solar wind ions (proton and alpha) are primarily in the field of view of SPAN-I, typically only a couple of days during each Encounter. The selection is done visually as shown in Fig. \ref{plasma_properties}, which is a good example of the core distributions being within the field of view of SPAN-I (i.e., the peak in the flux-angle spectrogram, top two panels, is fully resolved and not along the top or bottom of the field of view). 
 Proton and alpha speeds determined from moments are shown in the third panel of Figure \ref{plasma_properties}. This particular period is comprised of slow solar wind with the proton speed mostly exceeding that of the alpha particles. The bottom three panels show the radial magnetic field, $B_R$, the local Alfv\'en speed, $V_A$,  and the differential velocity over Alfv\'en speed ($|V_{\alpha p}|/V_A$), respectively. 
 The alpha-proton differential speed is lower than the local Alfv\'en speed for such a slow solar wind as previous missions observed at greater distances from the Sun \citep{Marsch_etal_1982_He, Neugebauer_etal_1996}. 
Previous studies have suggested that the Alfv\'en speed limits the alpha-proton differential velocity due to wave-particle interactions driven by alpha-proton instabilities \citep{Gary_etal_2000_instablity, Kasper_etal_2008_he}.
 Once the alpha-to-proton drift exceeds the $V_A$, it excites waves and instabilities in the plasma, such as the fast-magnetosonic/whistler instability \citep{Gary_etal_2000_instablity, Li_Habbal_2000_instability} and the Alfvén/ion-cyclotron instability \citep{Verscharen_etal_2013_AIC}. The generated waves interact with particles to satisfy the instabilities' threshold and thus limit  $V_{\alpha p}$ by its upper bound (i.e., $V_{\alpha p} < V_A$).

\begin{figure*}
\gridline{\fig{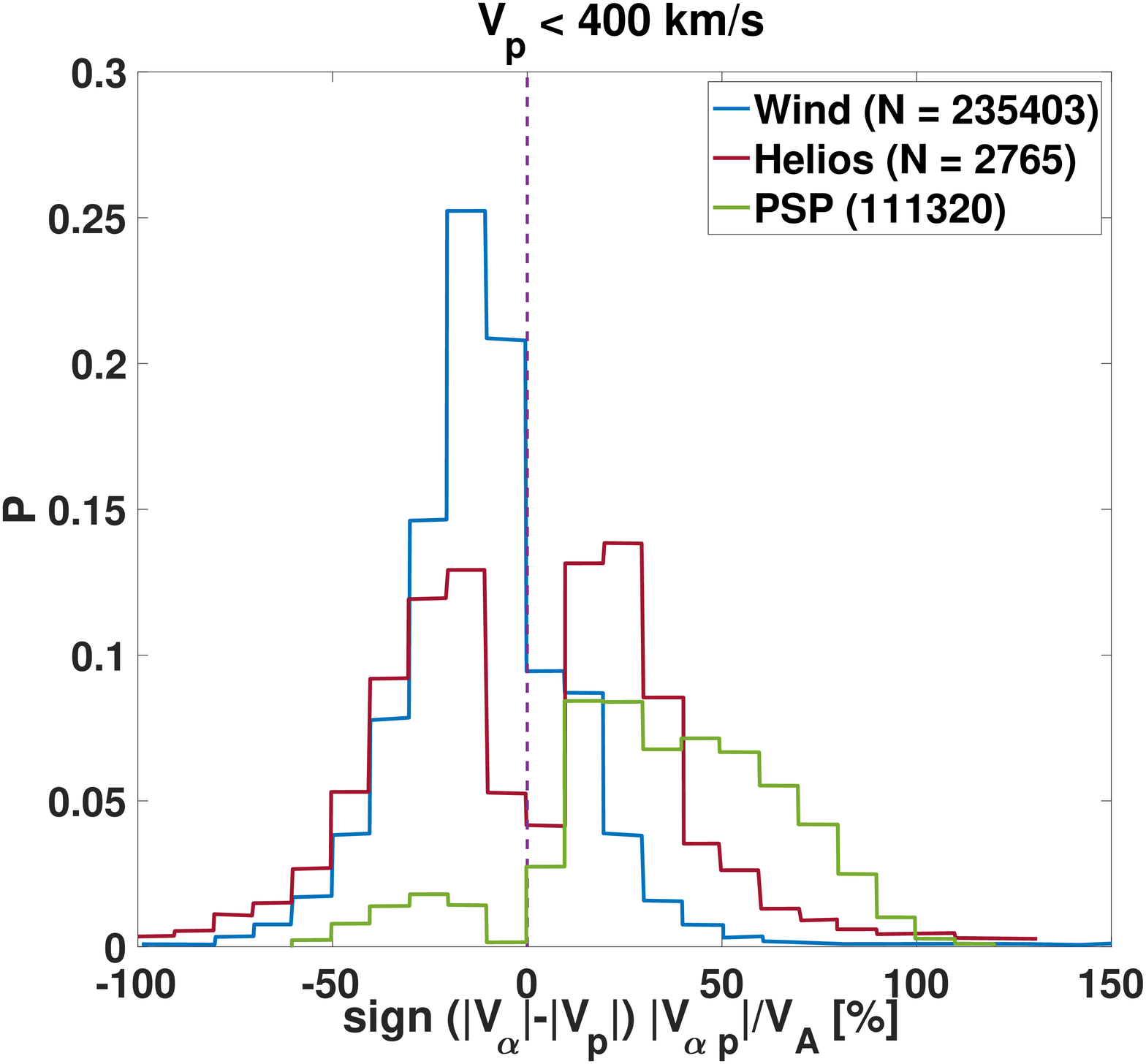}{0.3\textwidth}{(a)}
          \fig{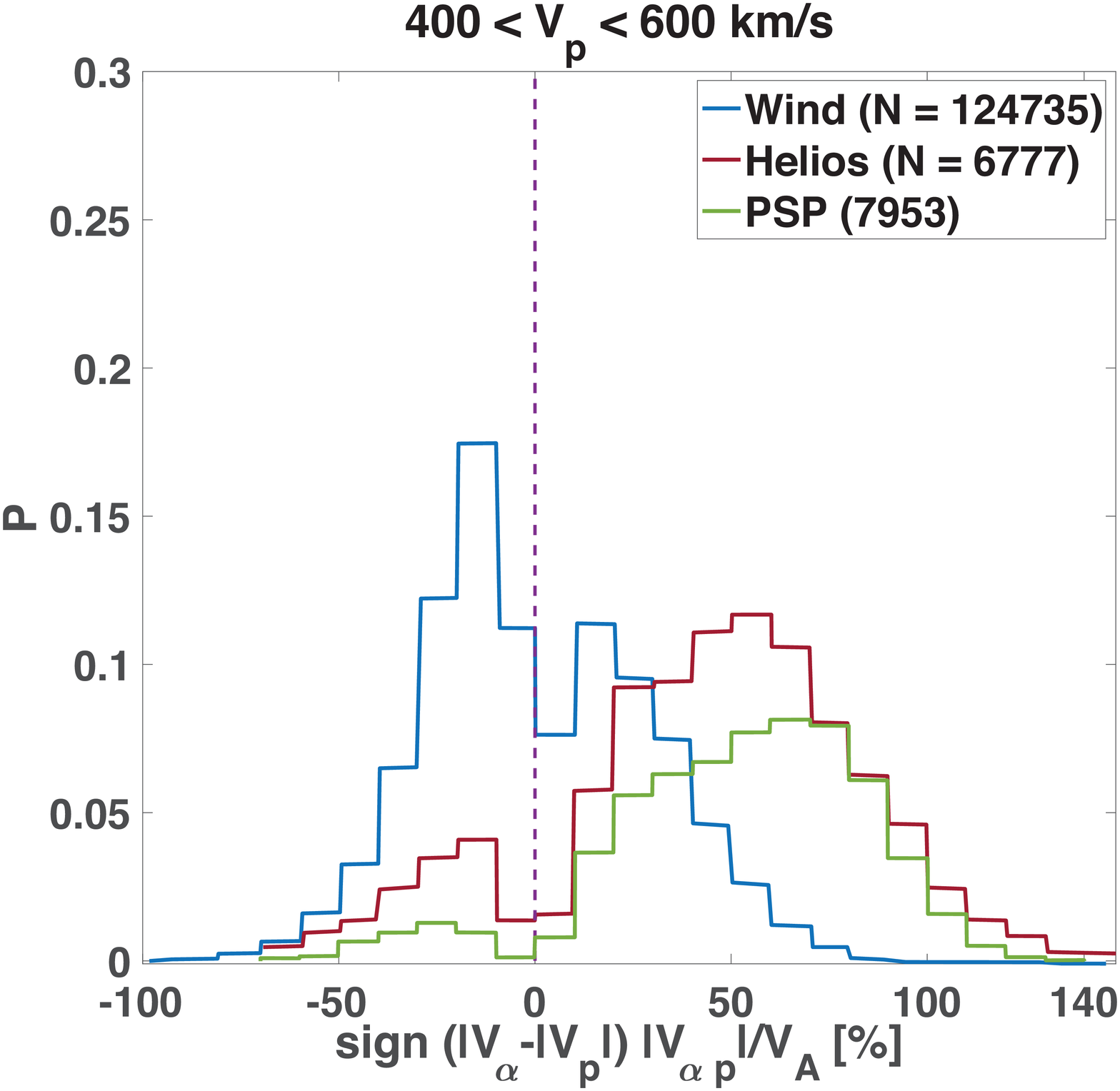}{0.3\textwidth}{(b)}
          \fig{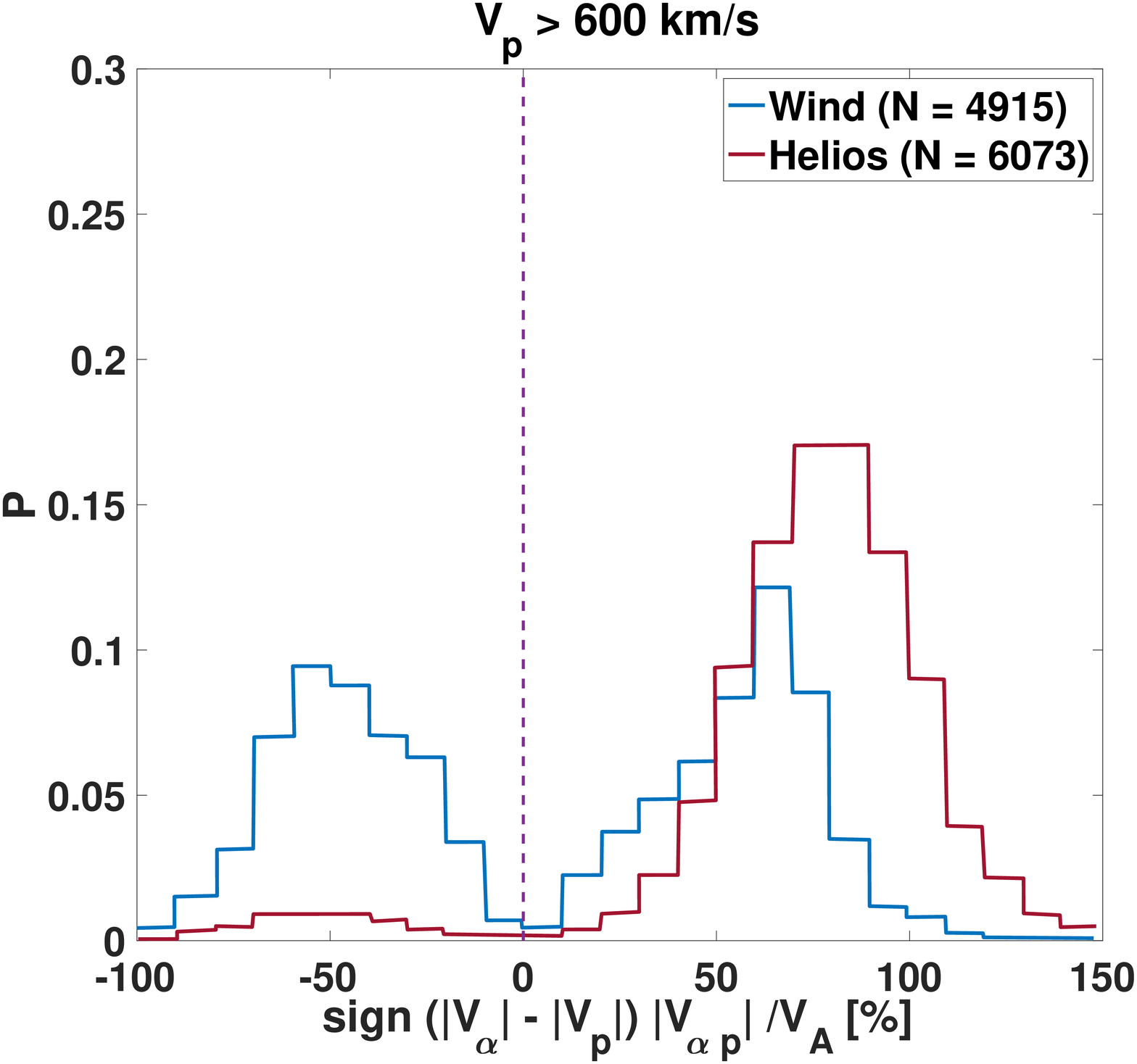}{0.3\textwidth}{(c)}
          }
\caption{Probability distributions of the alpha-proton differential speed normalized to Alfv\'en speed for slow to fast solar wind speeds (from left to right) are plotted. Wind data at 1 au (blue curves) and Helios data between 0.3 and 0.7 au (red curves) are adapted from Drovcova et al. (2017) and Marsch et al. (1982), respectively.  Green curves are PSP observations close to the Sun at about 0.1 au. N is the number of data set used for each analysis.
\label{Compare_3Spacecraft}}
\end{figure*}

Figure \ref{Compare_3Spacecraft} compares young solar wind data at about 0.1 au from PSP (green curves) with  Helios observations from 0.3 to 0.7 au \citep[red curves from][]{Marsch_etal_1982_He} and Wind measurements at 1 au \citep[blue curves from ][]{Durovcov_etal_2017_He}. Each panel shows the probability distributions of alpha-proton differential speed normalized to $V_A$ for different solar wind types. In comparison with previous spacecraft data, we divided our data set into three different categories, slow ($V_p < 400$ km/s), intermediate ($400 < V_p < 600$ km/s), and high ($V_p > 600$ km/s) speed solar winds.   
Note that negative values of normalized alpha-proton differential speed correspond to protons that travel faster than alpha particles, and positive values mean alpha particles are faster than protons. 
The left panel representing the slow solar wind shows a preferred positive $V_{\alpha p}$ in PSP data. Helios observations have two nearly identical positive and negative peaks. At larger distances, Wind observations at 1 au show only one peak with negative $V_{\alpha p}$. 
The prevalence of faster alphas compared to protons in the PSP observations close to the Sun confirms the super-acceleration or preferential acceleration of the alphas in the solar corona \citep{Ryan_Axford_1975_He, Marsch_etal_1982_He}. 
The number of negative $V_{\alpha p}$ observations increases as a function of distance from the Sun, as can be seen by comparing PSP to Helios and Wind observations. This pattern strongly supports the alphas' deceleration over distance during solar wind expansion  \citep{Neugebauer_etal_1994, Maneva_etal_2015}.  
The intermediate solar wind speed is shown in the middle panel. Wind observations still have a preferred negative $V_{\alpha p}$ while both PSP and Helios observations show approximately similar shapes with preferred positive peaks. The likely reason for not seeing a significant shift in PSP data to higher positive normalized differential velocities is that most of the immediate solar wind periods sampled are close to  400 km/s, and only contain a few periods with solar winds with speeds greater than 500 km/s. 
The last panel on the right shows the fast solar wind observed by Wind and Helios, however,  PSP did not observe high-speed solar wind during the intervals of Encounters 3-7 when the core of the solar wind was within the field of view of SPAN-I. As solar activity increases and PSP gets more connected to relatively large coronal holes, we expect to see high-speed solar wind (likely being characterized by positive $V_{\alpha p}$ and almost no negative values). 
In general, the comparison of PSP data with Helios and Wind shows an approximate evolution pattern of young solar wind close to the Sun to 1 au. 
Note that the data from three spacecraft (Wind, Helios, and PSP) are measured during different solar minimums.  

Furthermore, Figure \ref{Compare_3Spacecraft} shows that the $V_{\alpha p}/V_A$ ratio is generally less than one for different solar wind velocities at 1 au based on Wind observations.  Besides,  the Helios observations show that the flow difference between protons and alpha particles stays lower than or comparable to the $V_A$ for most solar wind speeds. With our data set, which includes up to Encounter 7, PSP has not seen many intervals with differential speeds greater than $V_A$.

\begin{figure}
\plotone{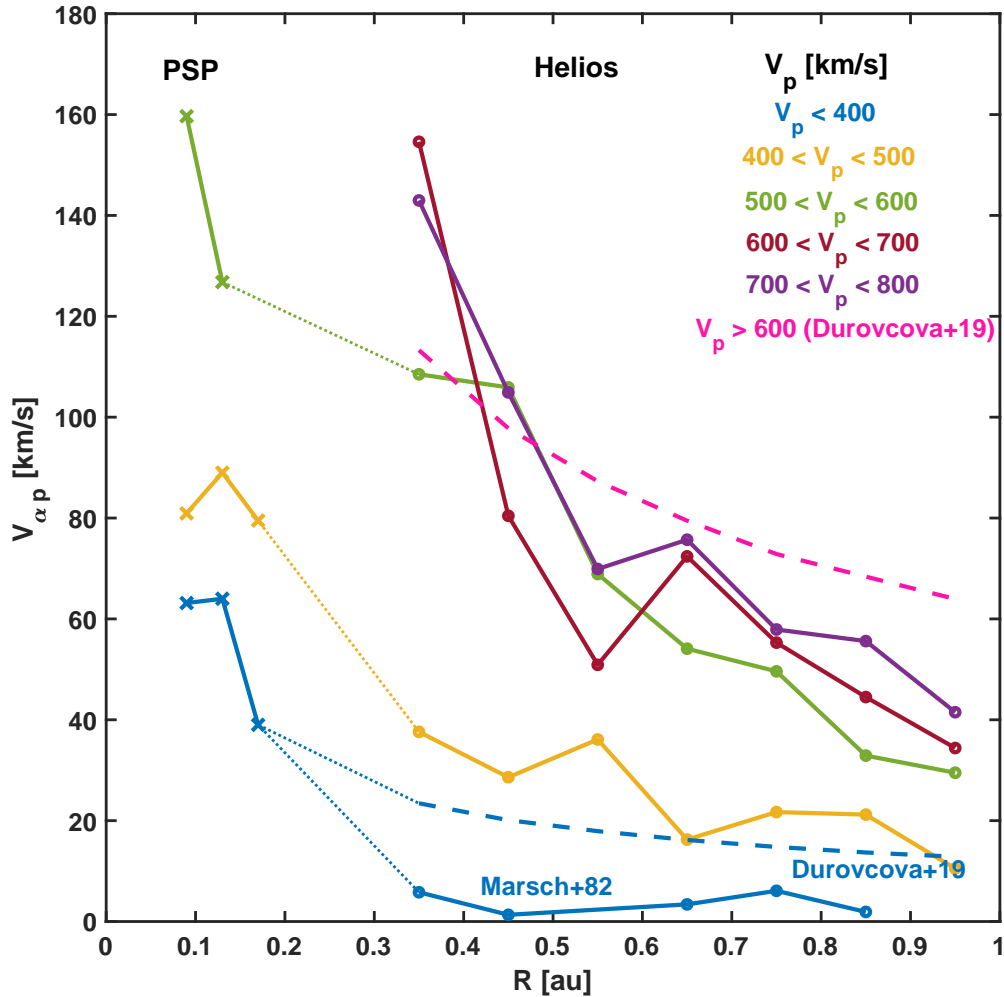}
\caption{Averages of the alpha-proton differential flow of Helios and PSP data sets as a function of distance from the Sun for different ranges of solar wind speeds are shown. Helios data are from 0.3 to 1 au. Circle points and dashed lines are reproduced from \cite{Marsch_etal_1982_He}  and \cite{Durovcova_etal_2019_Helios} fits, respectively. PSP data is shown with X, and the dotted lines connect the different data sets.
\label{different_R}}
\end{figure}

The dependence of the differential speed on both solar wind speed and distance from the Sun is apparent in Figure \ref{different_R}. Data points from 0.3 to 1 au are reproduced from Helios observations, in which circles and dashed lines are from \cite{Marsch_etal_1982_He} and \cite{Durovcova_etal_2019_Helios}, respectively. 
The ``X'' data points inside 0.2 au on the left side of the Figure \ref{different_R} are collected by PSP close to the Sun. 
Figure \ref{different_R} shows that PSP has only observed solar wind in the first three slow solar wind categories during our data sets of Encounters 3-7. Additionally, there was no observed solar wind with $V_p$ between 500 and 600 km/s during perihelia 3 when PSP was located at 35 $R_s$.   Generally, this figure shows that as solar wind becomes faster at a fixed distance from the Sun, its differential speed with alpha particles increases accordingly. 
The solar wind speed dependence of the differential speed is clearly seen from Helios data \citep{Marsch_etal_1982_He, Durovcova_etal_2019_Helios}. 
Young solar wind observations made by PSP are consistent with this dependency. The dotted lines in Figure \ref{different_R} connect the various data sets and to roughly estimate the change in differential speed between $\sim$0.15 to 0.3 au.
Generally, the radial dependence of the differential velocity is discernible from observations between $\sim $0.1 to 1 au, as $V_{\alpha p}$ decreases as solar wind propagates from the Sun to more considerable distances.  One possible explanation is that since the $V_A$ decreases with increasing distance, instabilities continuously decelerate alpha particles \citep{Gary_etal_2000_instablity} and thus their differential speed. Another possible reason is Coulomb collisions, which are found to be an essential mechanism to bring all species closer to thermal equilibrium at great distances from the sun \citep{Marsch_etal_1982_He, Kasper_etal_2008_he}. 
\cite{Marsch_etal_1982_He}, who analyzed  Helios data during a limited time of a solar minimum, observed an almost constant $V_{\alpha p}$ for slow solar wind with $V_p < 400$ km/s and thus suggested no radial dependence of differential velocity for slow solar winds. In another work, \cite{Durovcova_etal_2019_Helios} reanalyzed both Helios 1 and 2 data over an extended period during the whole solar cycle and showed a slight dependency of $V_{\alpha p}$ on the radial distance for the slow solar wind. 
PSP observations report that differential speed increases with decreasing heliocentric distance, even for slow solar wind during a solar minimum.
The significant radial dependence of $V_{\alpha p}$ was not as evident from the Helios data, which was farther away, as the slow ions were greatly thermalized before reaching the orbit of Helios.  

\cite{Kasper_etal_2017_preferential_heating} reported that preferential heating of the heavy ions is predominately active within some boundary near the Sun. Later, \cite{Kasper_etal_2019_preferential_heating} proposed that the boundary was well correlated to the the solar Alfv\'en critical surface (i.e, at about 20-40 $R_s$).  
PSP crossed the Alfv\'en surface for the first time on 28 April 2021 during its eighth encounter for about 5 hours \citep{Kasper_etal_2021_alfven_surface}. It will likely cross the Alfv\'en surface in future Encounters as well and thus will be able to measure the solar wind in this unexplored region directly.
Subsequent PSP Encounters would be critical to understanding the solar wind heating mechanism and the reason for preferential heating and acceleration of the alpha particles. 
Based on \cite{Kasper_etal_2019_preferential_heating} observation and prediction, the $V_{\alpha p}$ should not increase significantly after crossing the Alfv\'en point. However, if the super-acceleration of alphas happens near the solar surface, then $V_{\alpha p}$ should keep increasing during future PSP's data observations.

\begin{figure}
\plotone{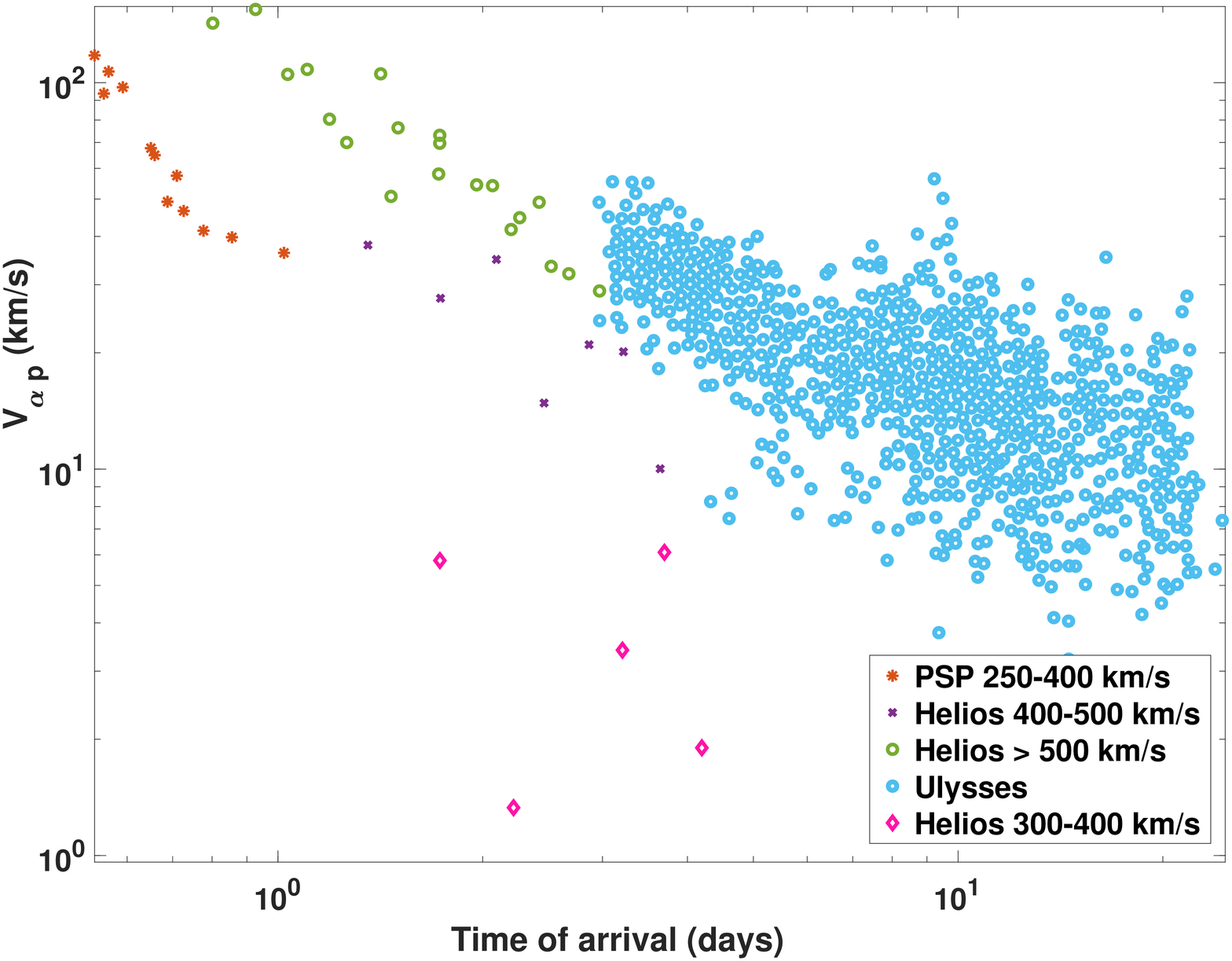}
\caption{Six-hour averages of alpha-proton differential velocity are shown as a function of time of arrival to the spacecraft in a log-log scale. Helios data and Ulysses data are adapted from \cite{Marsch_etal_1982_He} and \cite{Neugebauer_etal_1996}, respectively. 
\label{travel_time}}
\end{figure}

To better investigate how the alpha-proton differential velocity changes as the solar wind propagates away from the Sun, Figure 4 shows $V_{\alpha p}$ as a function of the time of arrival from the Sun to the point of observation by PSP, Helios, and Ulysses. 
Ulysses data are six-hour averages of the differential speed during the first five years of the mission (from 1991 to 1996) covering a large range of solar activities at heliographic latitudes from $+ 80^\circ$ to $- 80^\circ$ and radial distances from 1.3 to 5.4 au \citep{Neugebauer_etal_1996}.
Thus, the spread of observed $V_{\alpha p}$ in the Ulysses data comes from the various solar wind velocities and sources. 
The time of arrival of Helios data is computed from \cite{Marsch_etal_1982_He}.  
The purpose of plotting various Helios data speeds with different colors is to distinguish different solar wind kinds. It is clearly obvious that faster wind has larger $V_{\alpha p}$ which may suggest they have different source than slow winds.
The lowest-speed of Helios data with $V_p = 300 - 400$ km/s follows roughly a power law with PSP data, which its six-hour averages only includes solar wind with $V_p = 250 - 400$ km/s.  Once again, this plot confirms the dependence of the differential flow on travel time and distance from the Sun.  
Slow solar wind observed by PSP have a significant differential flow with alpha particles, which decreases as it propagates into the heliosphere reaching the orbit of Helios. 
As pointed out earlier, these observations are from different solar cycles, and spacecraft do not measure the identical solar wind. PSP mission Encounters have thus far measured limited statistics, mainly characterized by slow solar wind. Additional orbits of PSP, as we exit solar minimum with more available statistics, may provide faster speed solar wind with more statistics to better compare with Helios observations.

\section{Conclusion}
Our statistical analysis of PSP observations during Encounters 3-7 confirmed that the alpha-proton differential velocity is positively correlated with the bulk solar wind speed.
Thus,  differential flow between these two-particle populations may provide clues to the source of the solar winds and their acceleration and heating mechanism.
We showed that the dependence of differential flow on heliocentric distance  continued inside 0.3 au as  observed by PSP. Furthermore, PSP clearly showed that this dependency is also valid for slow solar wind, and they have larger alpha-proton differential speeds close to the Sun before their thermalizations ultimately occur. 
Close to the Sun, alpha particles are typically faster than protons showing their preferential acceleration mechanism. 
Our data set, which includes primarily slow solar wind, confirms that the magnitude of the alpha-proton differential speed is mainly below the local Alfv\'en speed.
While differential speed is oppositely correlated with the age of the solar wind, there is also a clear indication of different speed solar wind originating with various differential streaming.
Whether this difference between fast and slow differential flows comes from distinct sources or some processes such as collisions as the solar wind advects outwards remains an open question. Forthcoming studies will aim to focus on the critical relationship between the Coulomb collisions or collisional age and proton-alpha differential speed. 
Moreover, we will analyze future PSP Encounters at lower heliocentric distances, which may aid in our understanding of the super-acceleration process of alpha particles close to the Sun. This will likely help answer the first science objective of PSP by explaining how the solar wind is heated and accelerated.

\begin{acknowledgments}
Parker Solar Probe was designed, built, and is now operated by the Johns Hopkins Applied Physics Laboratory as part of NASA’s Living with a Star (LWS) program (contract NNN06AA01C). Support from the LWS management and technical team has played a critical role in the success of the Parker Solar Probe mission. 
The authors would like to thank J. Verniero, R. Livi, and A. Rahmati for useful discussions. The authors acknowledge CNES (Centre National d Etudes Spatiales), CNRS (Centre National de la Recherche Scientifique), the Observatoire de PARIS, NASA and the FIELDS/RFS team for their support to the PSP/SQTN data production, and the CDPP (Centre de Donnees de la Physique des Plasmas) for their archiving and provision. The FIELDS experiment on the PSP spacecraft was designed and developed under NASA contract NNN06AA01C. Finally, we thank the reviewer for the constructive comments that helped to improve our paper.
\end{acknowledgments}

%








\end{document}